# Giant negative magnetoresistance induced by the chiral anomaly in individual $Cd_3As_2$ nanowires

Cai-Zhen Li[1,†], Li-Xian Wang[1,†], Haiwen Liu[2], Jian Wang[2,3], Zhi-Min Liao[1,3*] & Da-Peng Yu[1,3*]

[1] State Key Laboratory for Mesoscopic Physics, Department of Physics, Peking University, Beijing 100871, China

[2] International Center for Quantum Materials, School of Physics, Peking University, Beijing 100871, China

[3] Collaborative Innovation Center of Quantum Matter, Beijing, China

†These authors contributed equally to this work.

*email: liaozm@pku.edu.cn; yudp@pku.edu.cn

**$Cd_3As_2$ is a newly booming Dirac semimetal with linear dispersion along all three momentum directions and can be viewed as 3D analog of graphene. As breaking of either time reversal symmetry or spatial inversion symmetry, the Dirac semimetal is believed to transform into Weyl semimetal with exotic chiral anomaly effect, while the experimental evidence of the chiral anomaly is still missing in $Cd_3As_2$. Here we report the magneto-transport properties of individual $Cd_3As_2$ nanowires. Large negative magnetoresistance (MR) with magnitude of -63% at 60 K and -11% at 300 K are observed when the magnetic field is parallel with the electric field direction, giving the evidence of the chiral magnetic effect in $Cd_3As_2$ nanowires. In addition, the critical magnetic field $B_C$, where there is an extremum of the negative MR, increases with increasing temperature. As the first observation of chiral anomaly induced negative MR in $Cd_3As_2$ nanowires, it may offer valuable insights for low dimensional physics in Dirac semimetals.**

Dirac materials such as graphene and topological insulators have attracted great attention recently, which possess relativistic and massless Dirac fermions with linear energy dispersion in two-dimensional momentum space[1-3]. As extending to three-dimensional (3D) space, the linear energy bands along all three directions in momentum space cross at crystal symmetry protected Dirac points, which is called as topological Dirac semimetals (TDSs)[4,5]. Recently, it is predicted that $A_3Bi$[6-9] (A=Na, K, Rb) and $Cd_3As_2$ [10-18] are candidates of 3D TDS materials in theory. Thereafter, the unique energy band structures have been identified by angle resolved photoemission spectroscopy (ARPES)[16-18] and scanning tunneling microscope (STM)[14] experiments in $Cd_3As_2$. Moreover, transport measurements[10,12,13,15] of $Cd_3As_2$ bulk crystals exhibit large linear magnetoresistance (MR), non-trivial quantum oscillations and Landau level (LL) splitting under magnetic field, which confirm the existence of 3D Dirac semimetal phase in $Cd_3As_2$. The Dirac point in 3D Dirac semimetal is composed of two overlapping Weyl nodes with opposite chirality, which can be separated in momentum space by breaking time-reversal symmetry or spatial inversion symmetry[11]. Applying a magnetic field can break up the time-reversal symmetry, resulting in the Dirac semimetals transform to Weyl semimetals. The splitting Weyl nodes are distributed along the external magnetic field and the distance between two Weyl nodes is proportional to the magnitude of the magnetic field[19]. An important feature of Weyl semimetal is the chiral anomaly effect[20-23], where the charges at the two splitting Weyl nodes are with opposite chirality. In the presence of magnetic field and electric field parallel with each other, the Weyl fermions residing at one Weyl node transport to another Weyl node with opposite chirality, resulting in non-conserved chiral charge and negative MR in transport measurements. Inspiringly, chiral anomaly induced negative MR has been observed in systems of $Bi_{1-x}Sb_x$[24], TaAs[25,26], $ZrTe_5$[27] and $Na_3Bi$[28]. Nevertheless, the negative MR is still not observed in $Cd_3As_2$, a widely investigated Dirac semimetal system. So far, the missing of negative MR in $Cd_3As_2$ causes an important issue that if the $Cd_3As_2$ system is exceptional or the theoretical prediction is not complete.

Here we employ the single crystal $Cd_3As_2$ nanowire with low carrier concentration to study the chiral anomaly effect on the transport properties. When magnetic field is perpendicular to electric field, a large positive MR up to 2000% is observed. While as magnetic field is parallel to electric field, a notable negative magnetoresistance is observed with magnitude of -63% at 60 K and -11% at 300 K, giving a clear evidence for chiral anomaly.

## Results

**Microstructures of the nanowire.** $Cd_3As_2$ is a group II-V compound and was intensively studied for its ultrahigh carrier mobility. It belongs to body centered tetragonal crystal system with $I_{41}cd$ space group. Here, we synthesized the $Cd_3As_2$ nanowires *via* chemical vapor deposition (CVD) method. The nanowires have large aspect ratio and demonstrate great flexibility, as shown by the scanning electron microscope (SEM) image in **Fig. 1a**. The nanowire length can be up to several hundred microns and the diameter ranges from 50 nm to 500 nm. **Fig. 1b** shows the transmission electron microscope (TEM) image of a typical nanowire with diameter of ~ 100 nm. The energy-dispersive X-ray spectroscopy (EDS) acquired from nanowire in the TEM is shown in **Fig. 1c**, and further quasi-quantitative analysis indicates that the atomic ratio of Cd and As is well consistent with the stoichiometric composition of $Cd_3As_2$. The high resolution TEM (HRTEM) image shown in **Fig. 1d** demonstrates the single crystal nature of the nanowire. The 0.73 nm inter-planar spacing suggests the [112] growth direction of the nanowires.

**Temperature dependent resistance.** Individual $Cd_3As_2$ nanowires were fabricated to contact with Au electrodes on a Si substrate with 285 nm $SiO_2$ layer. The schematic diagram of the device with four-probe measurement configuration is shown in **Fig. 2a**. Inset in **Fig. 2b** shows the optical image of a typical $Cd_3As_2$ nanowire device. The geometric parameters were obtained by SEM and atomic force microscopy (AFM) measurements (Supplementary **Fig. S1**). The temperature dependence of resistivity shown in **Fig. 2b** demonstrates that the resistivity increases as the temperature decreases from 300 K down to 19 K. While below 19 K, the resistivity decreases slightly with decreasing temperature. As far as we know, it is very different with the metallic behavior of $Cd_3As_2$ bulk crystal and the insulating behavior of $Cd_3As_2$ thin film[29]. The difference in the resistance-temperature behavior between $Cd_3As_2$ nanowires and its bulk counterpart may be caused either by different current direction (along [112] direction for the present nanowire situation while perpendicular to the [112] direction for previous bulk crystal studies[10,13,17]) or by the single crystalline nature of the nanowires. Additionally, compared with previous bulk crystals, one advantage of the nanowires is that it can avoid the uncertainty of the measured resistance caused by multiple conducting channels and current leakage. Due to the gapless energy band of the semimetal, the holes in the valence band can be thermally activated as the system has low carrier concentration[28]. The carrier concentration is extracted from the transfer curve of a typical $Cd_3As_2$ nanowire field-effect transistor (Supplementary **Fig. S2**). The carrier concentration of $n = 2.23 \times 10^{17}/cm^3$ is obtained at 1.5 K, which corresponds to the Fermi wave vector $k_F = 0.0187\text{Å}^{-1} = 0.47k_D$, where $k_D$ is wave vector of the Dirac node. Such low carrier concentration may be due to the little defects or impurities of the nanowires. From 100 K to 300 K, the data are well fitted by the Arrhenius equation of $\rho \sim \exp(E_a/k_B T)$ (**Fig. 2c**), where $k_B$ is the Boltzmann constant and $T$ is the temperature. The activation energy of the holes is extracted to be $E_a =$20.3 meV. The downward of the resistivity can be attributed to the reduction of carrier scattering and the improvement of the carrier mobility at low temperatures.

**Large positive magnetoresistance under B⊥E.** The resistivity increases with applying a magnetic field (B) perpendicular to the substrate plane. The $\rho - T$ behaviors under different magnetic fields are shown in **Fig. 2d**. It is clear that there is a peak in each $\rho - T$ curve, and the peak shifts to higher temperatures as increasing the magnetic field. As the magnetic field can lift the Fermi energy considering the 3D Fermi ellipsoid, the activation energy of the holes in the valance band will increase with increasing magnetic field (Supplementary **Fig. S3**), resulting in a dramatic increase of the resistivity as temperature going down from 300 K under high magnetic field. The magnetoresistance (MR) defined as $MR = \frac{\rho(B)-\rho(0)}{\rho(0)} \times 100\%$ are presented in **Fig. 2e**. Under 14 T, the MR is 210% at 1.5 K, and reaches the maximum of 2140% at 80 K, then reduces to 226% at 300 K. The largest MR appeared at moderate temperature range is in accordance with the non-metallic $\rho - T$ behavior under magnetic field shown in **Fig. 2d**. On one hand, at high temperatures (T > 100 K), the Landau level at high magnetic field is thermally smeared and the magnetic field induced resistance change is relatively small compared with that at low temperatures (Supplementary **Fig. S4**). On the other hand, at low temperatures (T < 30 K), the zero-field resistance R(B=0) is very large. Since the MR = R(14 T)/R(0 T) -1, it is expected to observe the largest MR at moderate temperatures. To compare the MR behaviors at different temperatures, the MR curves at 1.5 K and 300 K are magnified 10 times and plotted with the MR at 80 K together in **Fig. 2f**. In the low magnetic field region, the MR at 1.5 K increases fast and can be attributed to the weak anti-localization effect, while the MR at 300 K shows a quadratic increase and is due to the classical theory of the carriers with Lorentz force. At 1.5 K, the MR demonstrates linear field dependence above 10 T, which is well consistent with previous report and can be understood by considering that the suppression of backscattering can be released by magnetic field[15]. At high temperatures, the MR grows dramatically under high magnetic field because the magnetic field enlarges the activation energy of the holes in the valance band.

**Negative MR under B//E.** To further identify the anisotropy of the MR, angular dependence of MR was performed by varying the relative orientation between the current direction and the magnetic field direction in the plane perpendicular to the substrate. As shown in the inset in **Fig. 3a**, the $\theta$ represents the angle between the magnetic field direction and normal direction of the substrate. To verify the experimental phenomena, we have measured about ten samples. Typical results are presented in **Fig. 3**. The results are well repeatable and similar results are obtained from different devices (Supplementary **Figs. S5-S6**). The MR decreases apparently as the direction of the magnetic field changes from $B \perp E$ ( $\theta = 0°$) to $B//E$ ( $\theta = 90°$), as shown in **Fig. 3a**. It is interesting to note that there is a negative MR as $B//E$. The low field behaviors of the MR under $B//E$ and at different temperatures are shown in **Fig. 3b**. At zero B, an obvious dip of the MR is exhibited at 1.5 K, and then gradually disappears with increasing temperature to 10 K. This MR dip is ascribed to the weak anti-localization effect as a result of strong spin-orbit interactions in $Cd_3As_2$. Nevertheless, the negative MR under low magnetic field is very robust against to temperature. In order to study the negative MR in more details, the MR curves at various temperatures under $B//E$ are given in **Fig. 3c**. The maximum magnitude of the negative MR is up to -63% at 60 K and 7 T. The negative MR ~ -11% is still observed at 300 K. The negative MR has an extremum at the critical magnetic field $B_C$, and then the magnitude decreases with increasing magnetic field, similar to that observed in Bi$_{1-x}$Sb$_x$[24] and TaAs[25,26] systems.

## Discussion

Because the negative MR is rather robust and survives at room temperature, the weak localization effect due to the quantum interference is not the origin of the observed negative MR. Here we attribute the observed negative MR to the chiral anomaly in Cd$_3$As$_2$ nanowires. The Dirac point described by four-component Dirac equations in Dirac semimetals is composed by two Weyl nodes with opposite chirality (right-handed or left-handed). Applying a magnetic field, these two Weyl nodes would be separated in momentum space along with the direction of the magnetic field[24,30]. Initially the right- and left-handed fermions have equal chemical potential $\mu^R = \mu^L$. In the presence of parallel electric field, there would be an imbalance ($\mu^R \neq \mu^L$) between two Weyl nodes with opposite chirality. In such a case, the continuity equation of right- or left-handed Weyl node takes the form of $\nabla \cdot j^{R,L} + \partial_t \rho^{R,L} = \pm \frac{e^3}{4\pi^2 \hbar^2 c} \vec{E} \cdot \vec{B}$[31]. The chiral charge at a single Weyl node is not conserved. That is the so-called chiral anomaly. The charge depleted at one Weyl node will be generated at the other node with opposite chirality (charge pumping) and thus the charge is conserved overall the system. There is a net current generation as a result of chiral imbalance in the form of $j_c = j_c^R - j_c^L = \frac{e^2 B}{4\pi^2 \hbar^2 c}(\mu^R - \mu^L)$, where $j_c$ is named as chiral current. Since the chiral current is in the direction of electric field, a negative MR will be induced. The difference of the chemical potential between two Weyl nodes with opposite chirality follows the relation of $\mu^R - \mu^L \propto \vec{E} \cdot \vec{B}$[24,27]. Therefore, the chiral current induced correction of conductance should have $\Delta\sigma \propto B^2$.

For the Cd$_3$As$_2$ system, there are a pair of Dirac points along k$_z$ direction and located at $\pm k_z^D$ near the high symmetric point Γ, as marked by red dots in **Fig. 4a**. The magnetic field applied in [112] direction splits the Dirac points into Weyl nodes along the direction of magnetic field, as marked by the green and blue dots in **Fig. 4a**. As applying an electric field, an additional chiral current is induced due to the different chemical potentials of the Weyl nodes (**Fig. 4b**). As shown in **Fig.4c**, the field dependence of the positive conductivity (negative MR) can be well fitted using the equation of $\sigma = \sigma_0 + C_a(T)B^2$, where $\sigma_0$ is the zero field conductivity.

It is worth to note that the MR curves at low temperatures have two minima. The first minimum point at about 3 T becomes indistinct as increasing temperature to 30 K. Due to the low carrier concentration, the upturned MR under magnetic field above 3 T may be due to the Landau level gap. As increasing temperature, more holes in the valance band can be thermally activated to over this gap, resulting in the $B_{c1}$ increases notably with increasing temperature (**Fig. 4d**). The second minimum point at about 7.5 T seems to be more robust and becomes to be dominant as temperature is higher than 20 K, as seen from the MR evolution with varying temperature in **Fig. 3c**. Under high magnetic field large than 7 T, the attenuation of the negative MR may be due to the Coulomb interaction among the electrons occupying the chiral states[25]. The critical magnetic field $B_{c2}$ has a weak temperature dependence below 100 K and then shifts to high magnetic field at higher temperature (**Fig. 4d**). At room temperature, the negative MR dominates the entire range from 0 to 14 T, suggesting that the chiral states are robust against thermal perturbation. These new exotic properties are comprehensive and need to be further studied.

In summary, we have studied the magnetotransport properties of individual Dirac semimetal $Cd_3As_2$ nanowires. Benefiting from the single crystalline nature and the low carrier concentration of the nanowire, giant negative MR of -63% at ~ 60 K is observed as the magnetic field is parallel to electric field. The negative MR is still notable with the value of -11% at 300 K. The observations give clear evidence for the chiral magnetic effect in $Cd_3As_2$ system, which provides an ideal platform to explore new physical phenomena of 3D Dirac and Weyl semimetals.

## Methods

**Growth and characterization of the nanowires.** The $Cd_3As_2$ nanowires were prepared by CVD method in tube furnace. $Cd_3As_2$ powders were placed at the center of the furnace and silicon wafers coated with a thin gold film about 5 nm in thickness were used as substrates to collect the products downstream. First the tube furnace was flushed several times with Argon gas to fully get rid of oxygen. Then the temperature was gradually increased to 650℃ and kept for 10 minutes during with Argon flow of 20 sccm as carrier gas. After the growth process, the furnace was cooled naturally. The SEM characterizations were performed in a FEI Nano430 SEM system, and the TEM characterizations were performed in a FEI Tecnai F20 TEM equipped with EDS system.

**Transport measurements.** The synthesized $Cd_3As_2$ nanowires were transferred to a silicon substrate with an oxide layer of 285 nm, and then electrodes were fabricated after a series of processes including electron beam lithography, deposition of metal electrodes and lift-off. All the magneto-transport measurements were carried out in a commercial Oxford system, which can offer magnetic field up to 14T and base temperature down to 1.5 K. Four-probe electrical measurements were performed through Stanford SR830 lock-in amplifiers by supplying 0.1 μA current with frequency of 17.7 Hz.

## Acknowledgements

This work was supported by MOST (Nos. 2013CB934600, 2013CB932602) and NSFC (Nos.

11274014, 11234001). We are grateful to Professor Xincheng Xie at PKU for inspired discussions.

## Author contributions

Z.-M.L. conceived and designed the experiments. D.-P.Y. gave scientific advice. C.-Z.L. performed the nanowire growth, characterization and device fabrication. L.-X.W. and C.-Z.L. performed the transport measurements. H.W.L. and J.W. contributed to analyze the data. Z.-M.L. and C.-Z.L. wrote the manuscript.

**Competing financial interests:** The authors declare no competing financial interests.

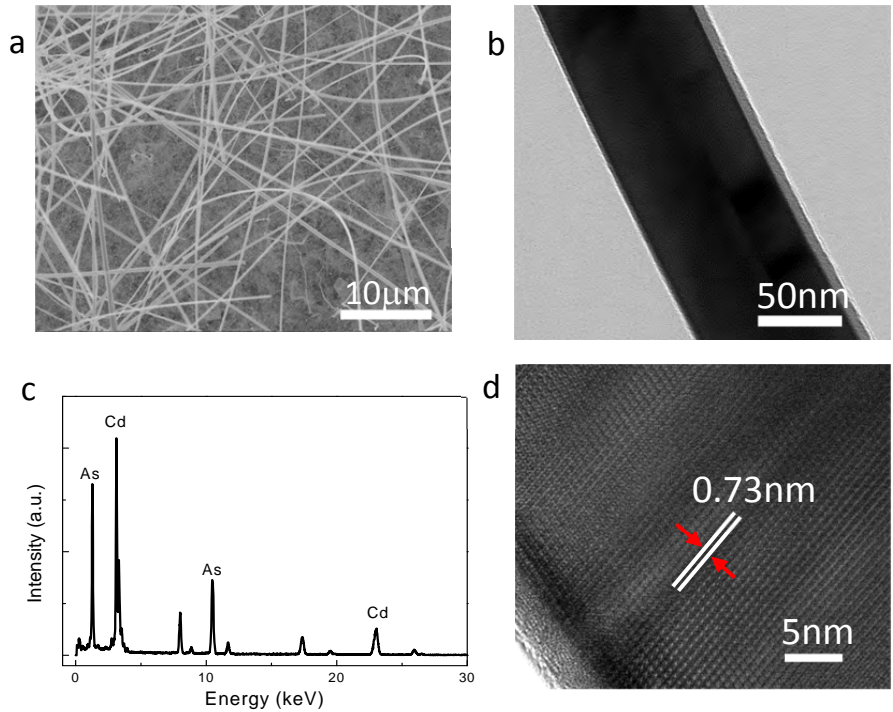

**Figure 1 | Characterization of synthesized Cd$_3$As$_2$ nanowires.** (a) SEM image of the nanowires. (b) TEM image, (c) the EDS spectrum, and (d) HRTEM image of a nanowire. The 0.73 nm interplanar spacing indicates the [112] growth direction.

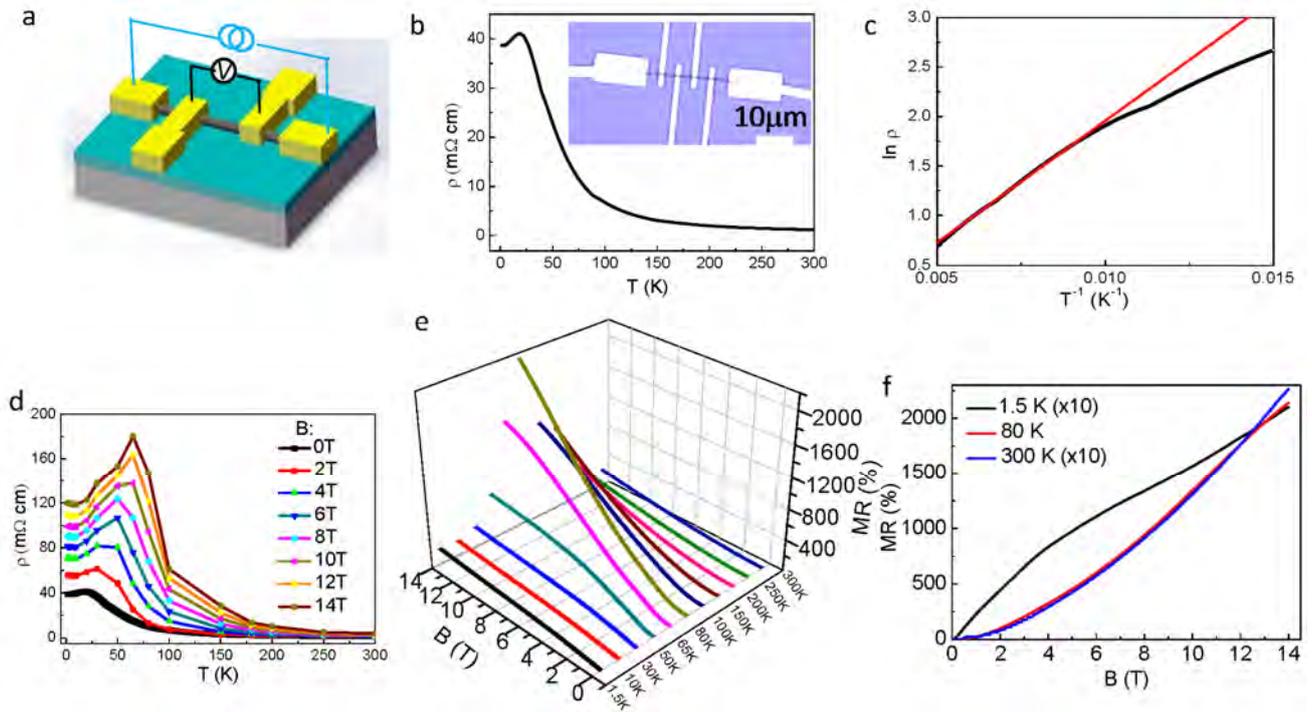

**Figure 2 | Temperature dependence of resistivity and MR under perpendicular B.** (a) Schematic diagram of the device. (b) Temperature dependence of resistivity. The inset in (b) is an optical image of a typical device. (c) Arrhenius plot of the $\rho \sim T$ curve. (d) The $\rho \sim T$ curves under different B. (e) The MR behaviors at different temperatures. (f) The MR at 1.5 K and 300 K are magnified 10 times and plotted with the MR at 80 K for comparison.

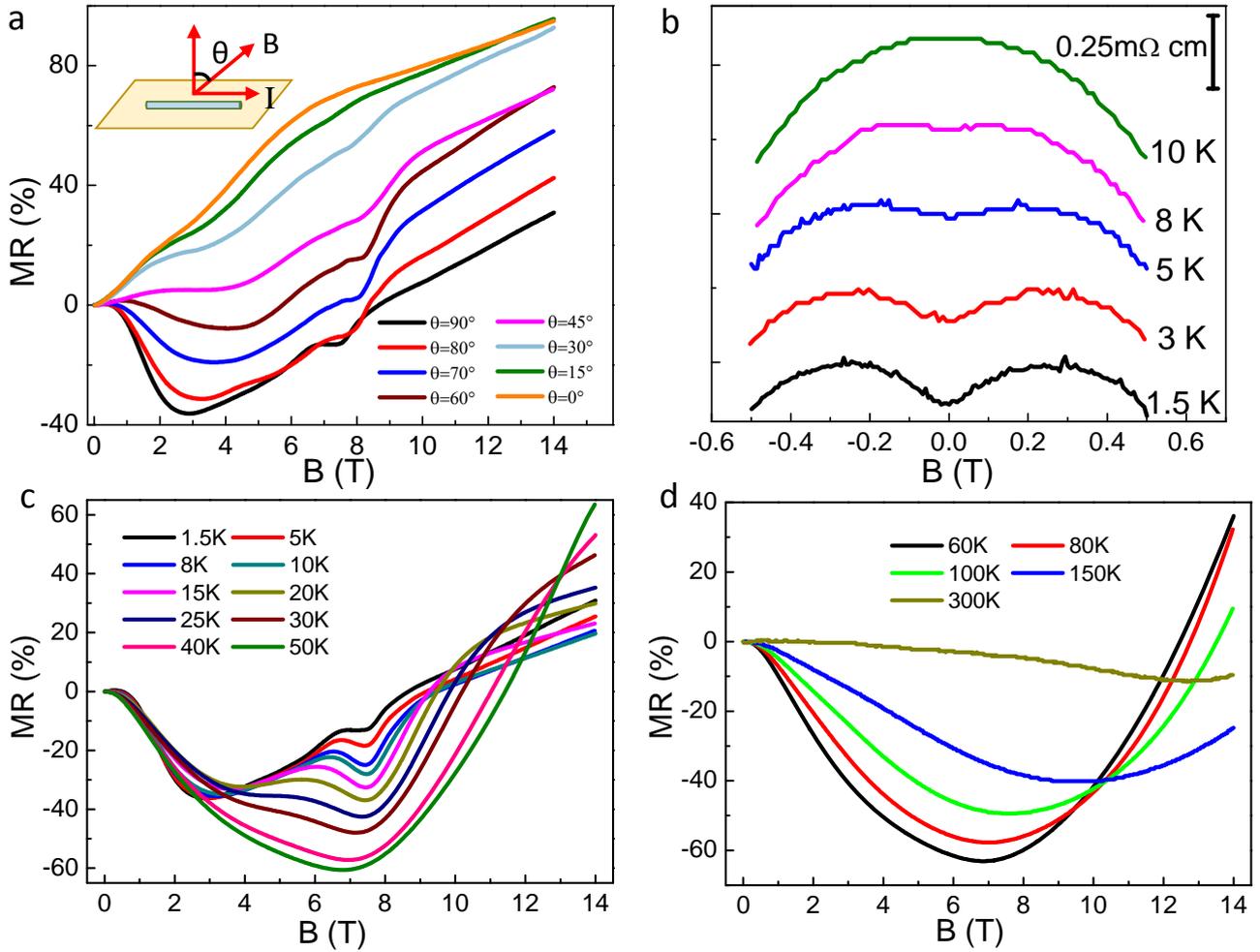

**Figure 3 | Negative MR under B//E.** (a) MR as varying the magnetic field from B⊥E ( θ = 0°) to B//E ( θ = 90°). (b) Field dependence of the resistivity under B//E and B < 0.5 T to show the weak anti-localization effect at low temperatures. (c) and (d) MR measured under B//E and temperatures from 1.5 K to 300 K.

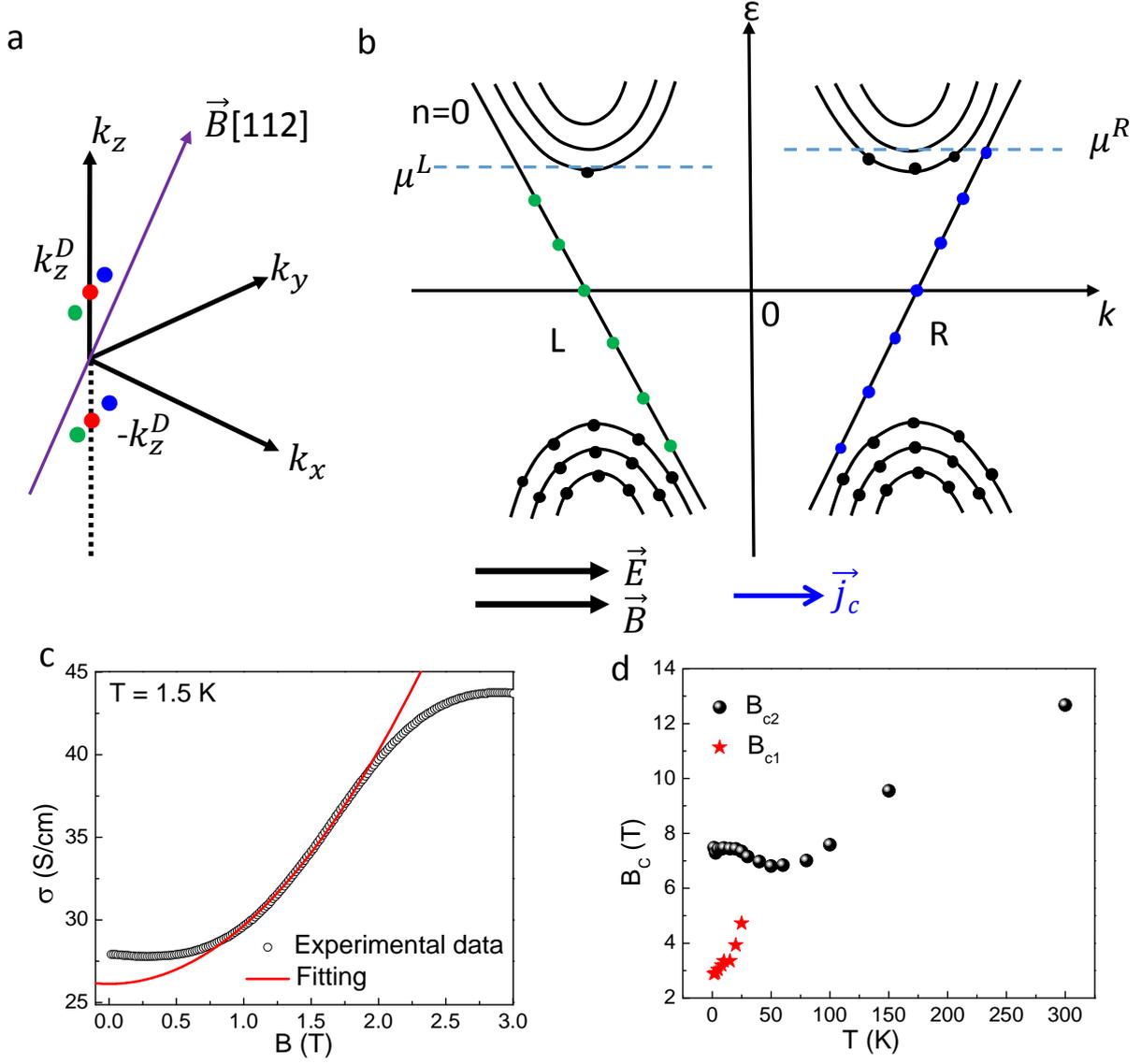

**Figure 4 | Mechanism of the negative MR.** (a) A pair of Dirac points (red dots) along the $k_z$ near the Γ point in k-space in Cd₃As₂ system. Applying a magnetic field along the [112] direction, the Dirac point splits into two Weyl nodes with opposite chirality (blue and green dots) along the field direction. (b) Schematic of the chiral anomaly induced current. The filled circles represent the occupied LLs. The states at lowest LL and at the two different Weyl nodes are with opposite chirality. The blue arrow denotes the net current along the field in case that B parallel to E. (c) Fitting result of the chiral anomaly induced positive conductivity (negative MR) at 1.5 K using $\sigma \propto B^2$ under low magnetic field. (d) The critical magnetic field as a function of temperature.

# Supplementary Figures

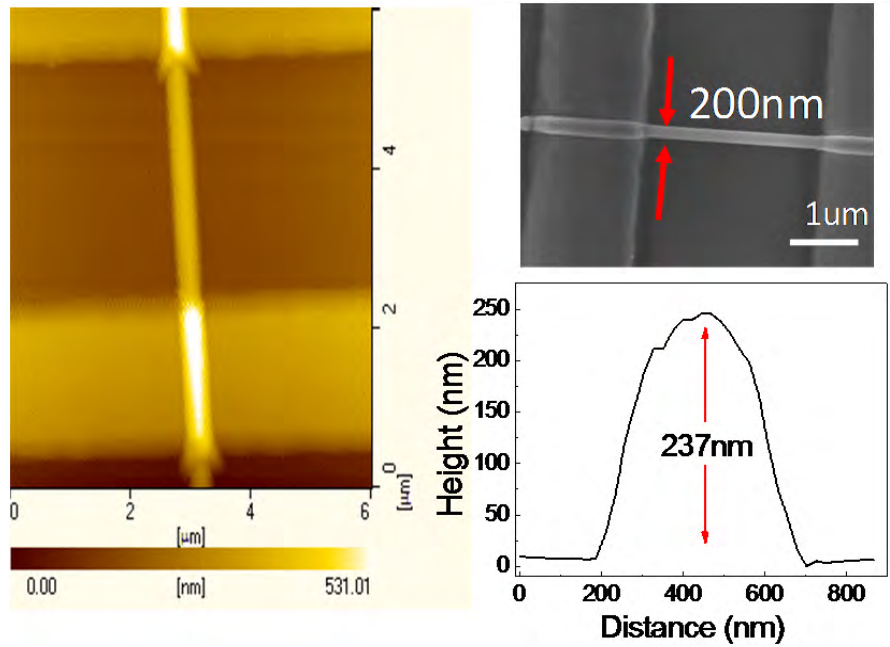

**Supplementary Figure S1.** AFM and SEM results of a typical device.

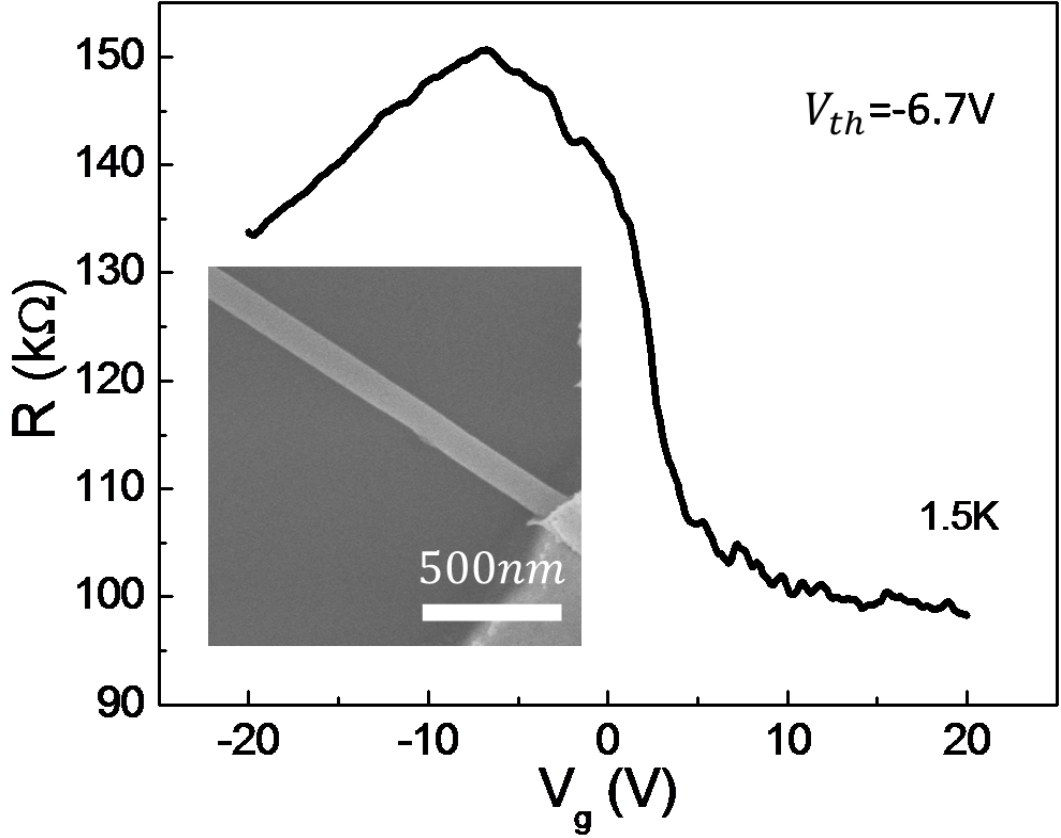

**Supplementary Figure S2.** Back gate voltage modulation of the resistance of a typical nanowire field effect transistor at 1.5 K. Inset: SEM image of the nanowire device. The carrier density can be calculated by $n = C \frac{1}{eS}(V_g - V_{th})$, where $V_g$ is the gate voltage, $V_{th}$ is the threshold voltage (charge neutral point), S is the cross-section area of the nanowire, and $C$ is the $SiO_2$-nanowire capacitance with $\frac{C}{l} = \frac{2\pi\varepsilon_0\varepsilon_r}{cosh^{-1}(\frac{r+h}{r})}$, $l$ is the length of the nanowire between two voltage probes, and $r$ is the radius of the nanowire, $h$ is the gate oxide thickness, and $\epsilon_r = 3.9$ is the relative dielectric constant of the oxide layer $SiO_2$. Here, $r$ = 50 nm, $l$ = 2 μm, and $h$ = 285 nm.

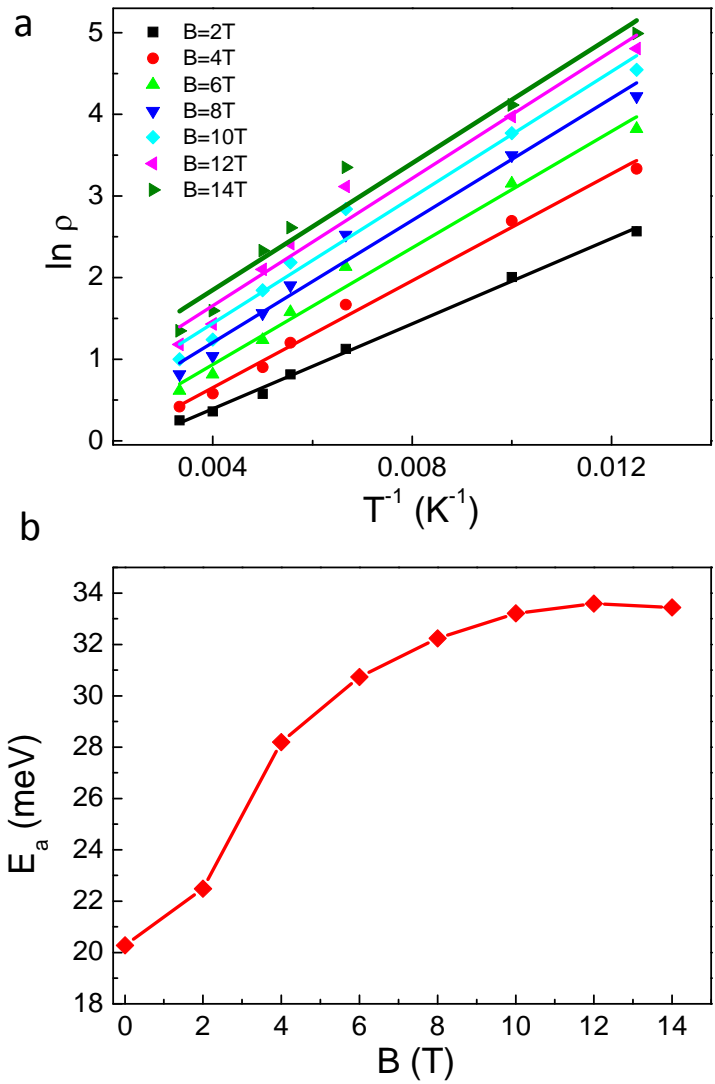

**Supplementary Figure S3.** (a) The Arrhenius plot from the curves in Figure 2d under different magnetic field. (b) The activation energy calculated under different magnetic field.

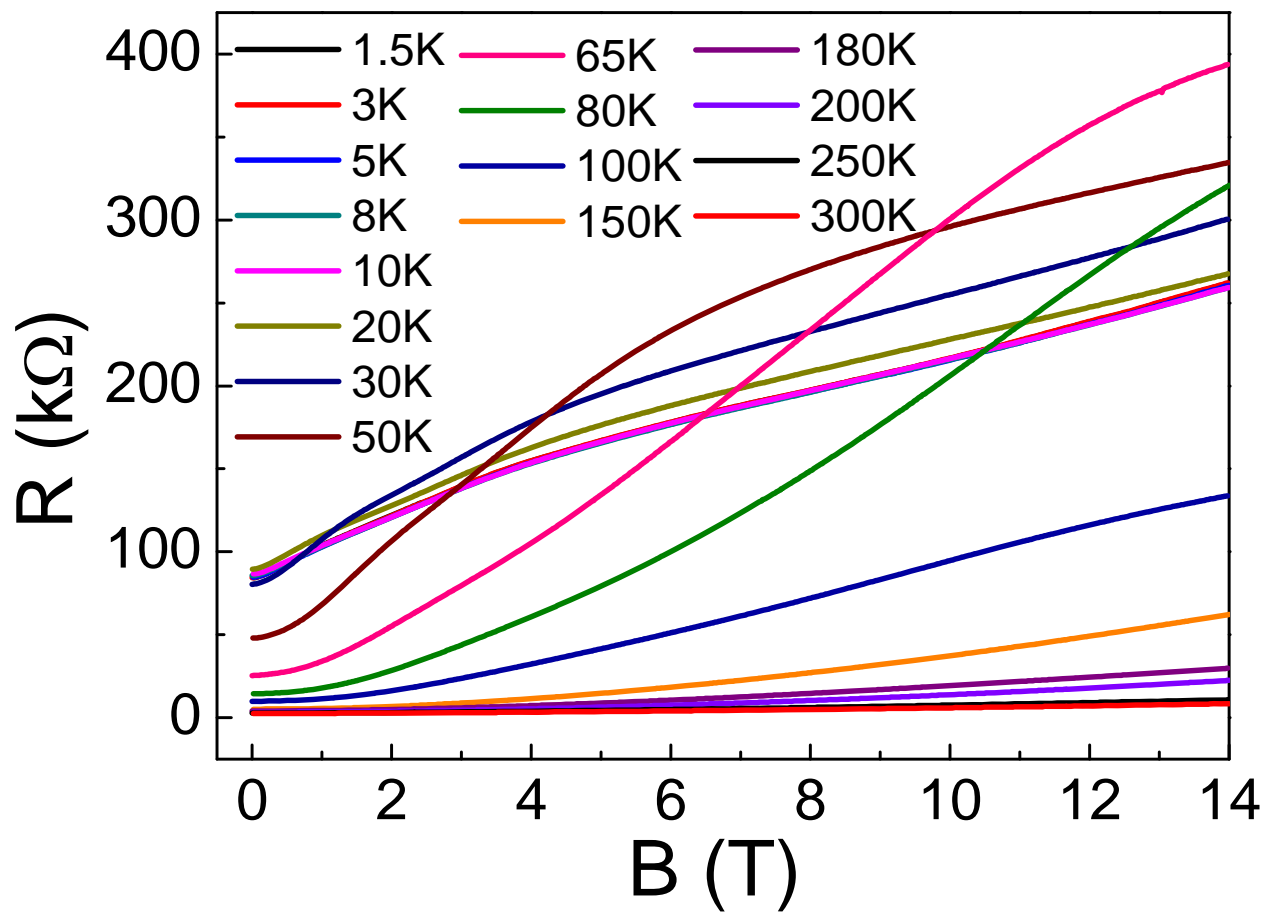

**Supplementary Figure S4.** The resistance as a function of B under perpendicular magnetic field and at different temperatures as denoted, which are the raw data of the MR shown in Figure 2e.

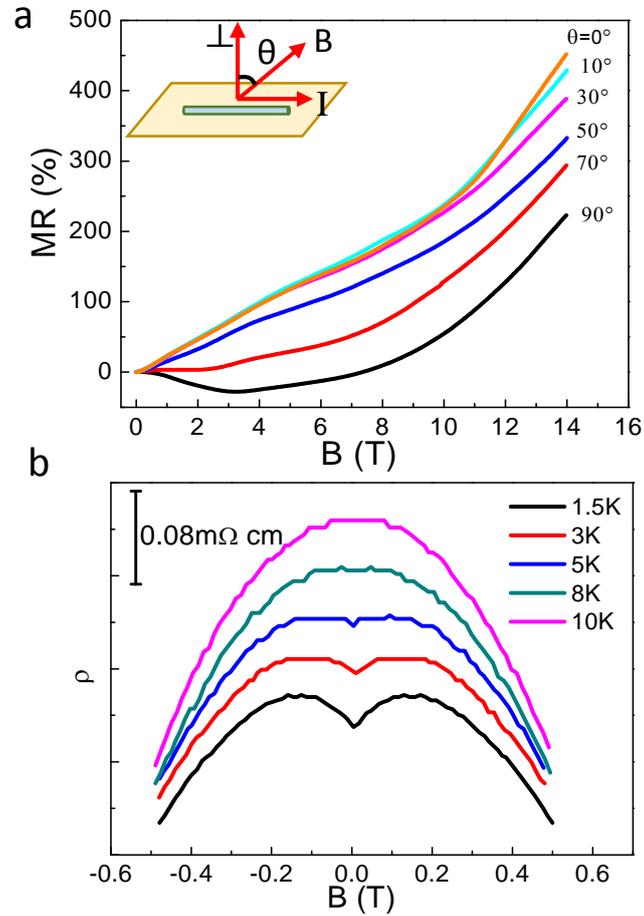

**Supplementary Figure S5.** Another similar device shows similar magnetotransport properties as that in Figures 3a and 3b. (a) MR as varying the magnetic field from B⊥E ( θ = 0°) to B//E ( θ = 90°). (b) Field dependence of the resistivity under B//E and B < 0.5 T. The weak anti-localization effect is clearly observed under low magnetic field and at low temperatures.

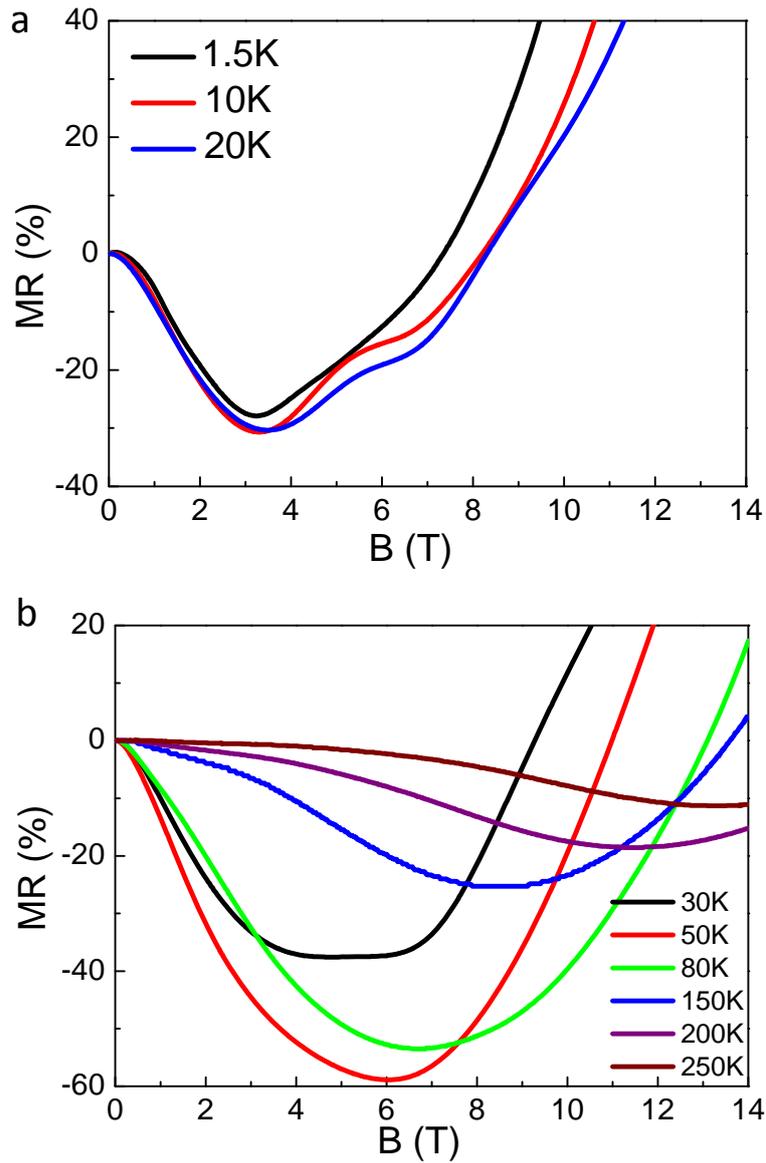

**Supplementary Figure S6.** Another similar device under B//E shows similar negative MR properties as that in Figures 3c and 3d. The MR curves are separated in two figures to clearly show the evolution of the negative MR as varying temperature.